\documentclass[11pt, a4paper]{iopart}
\pdfoutput=1

\usepackage{etoolbox}

\makeatletter
\long\def\@makefntext#1{\parindent 1em\noindent 
 \makebox[1em][l]{\footnotesize\rm$\m@th{{}^\arabic{footnote}}$}%
 \footnotesize\rm #1}
\def\@makefnmark{\hbox{${}^{\arabic{footnote}}\m@th$}}
\def\@thefnmark{\arabic{footnote}}
\makeatother

\usepackage{iopams}
\expandafter\let\csname equation*\endcsname\relax
\expandafter\let\csname endequation*\endcsname\relax
\usepackage{amsmath}

\usepackage[numbers,sort&compress]{natbib} 

\usepackage[T1]{fontenc}
\usepackage{ulem}
\usepackage{paralist}
\usepackage{graphicx}
\usepackage{amssymb}
\usepackage{latexsym}
\usepackage{fancyhdr}
\usepackage{array}
\usepackage{amsfonts}
\usepackage{mathrsfs}
\usepackage{mathtools}
\usepackage{color}
\usepackage{verbatim}
\usepackage{physics}
\usepackage[caption=false]{subfig}
\usepackage{enumitem}
\usepackage{bbold}
\usepackage[colorlinks=true,urlcolor=blue,citecolor=blue,linkcolor=blue]{hyperref}

\usepackage{float}
\usepackage{fullpage}

\newcommand{\avg}[1]{\langle #1\rangle}

\newcommand{\BE}{\begin{equation}}
\newcommand{\EE}{\end{equation}}

\newcommand{\A}{\mathbb{A}}
\newcommand{\X}{\mathbb{X}}
\newcommand{\Y}{\mathbb{Y}}
\newcommand{\Bo}{\mathbb{B}}
\newcommand{\pr}{\text{\textbf{P}}}
\newcommand{\rp}{\operatorname{\mathbb{R}e}}

\newcommand{\BFA}{\mathrm{I}}

\usepackage{xcolor}

\newcommand{\mk}[1]{{\color{black} #1}}

\newcommand{\blk}{\color{black}}

\providecommand{\DIFaddbegin}{} 
\providecommand{\DIFaddend}{} 
\providecommand{\DIFdelbegin}{} 
\providecommand{\DIFdelend}{} 
\providecommand{\DIFaddbeginFL}{} 
\providecommand{\DIFaddendFL}{} 
\providecommand{\DIFdelbeginFL}{} 
\providecommand{\DIFdelendFL}{} 
\newcommand{\DIFscaledelfig}{0.5}
\RequirePackage{settobox} 
\RequirePackage{letltxmacro} 
\newsavebox{\DIFdelgraphicsbox} 
\newlength{\DIFdelgraphicswidth} 
\newlength{\DIFdelgraphicsheight} 
\LetLtxMacro{\DIFOincludegraphics}{\includegraphics} 
\newcommand{\DIFaddincludegraphics}[2][]{{\color{blue}\fbox{\DIFOincludegraphics[#1]{#2}}}} 
\newcommand{\DIFdelincludegraphics}[2][]{
\sbox{\DIFdelgraphicsbox}{\DIFOincludegraphics[#1]{#2}}
\settoboxwidth{\DIFdelgraphicswidth}{\DIFdelgraphicsbox} 
\settoboxtotalheight{\DIFdelgraphicsheight}{\DIFdelgraphicsbox} 
\scalebox{\DIFscaledelfig}{
\parbox[b]{\DIFdelgraphicswidth}{\usebox{\DIFdelgraphicsbox}\\[-\baselineskip] \rule{\DIFdelgraphicswidth}{0em}}\llap{\resizebox{\DIFdelgraphicswidth}{\DIFdelgraphicsheight}{
\setlength{\unitlength}{\DIFdelgraphicswidth}
\begin{picture}(1,1)
\thicklines\linethickness{2pt} 
{\color[rgb]{1,0,0}\put(0,0){\framebox(1,1){}}}
{\color[rgb]{1,0,0}\put(0,0){\line( 1,1){1}}}
{\color[rgb]{1,0,0}\put(0,1){\line(1,-1){1}}}
\end{picture}
}\hspace*{3pt}}} 
} 
\LetLtxMacro{\DIFOaddbegin}{\DIFaddbegin} 
\LetLtxMacro{\DIFOaddend}{\DIFaddend} 
\LetLtxMacro{\DIFOdelbegin}{\DIFdelbegin} 
\LetLtxMacro{\DIFOdelend}{\DIFdelend} 
\DeclareRobustCommand{\DIFaddbegin}{\DIFOaddbegin \let\includegraphics\DIFaddincludegraphics} 
\DeclareRobustCommand{\DIFaddend}{\DIFOaddend \let\includegraphics\DIFOincludegraphics} 
\DeclareRobustCommand{\DIFdelbegin}{\DIFOdelbegin \let\includegraphics\DIFdelincludegraphics} 
\DeclareRobustCommand{\DIFdelend}{\DIFOaddend \let\includegraphics\DIFOincludegraphics} 
\LetLtxMacro{\DIFOaddbeginFL}{\DIFaddbeginFL} 
\LetLtxMacro{\DIFOaddendFL}{\DIFaddendFL} 
\LetLtxMacro{\DIFOdelbeginFL}{\DIFdelbeginFL} 
\LetLtxMacro{\DIFOdelendFL}{\DIFdelendFL} 
\DeclareRobustCommand{\DIFaddbeginFL}{\DIFOaddbeginFL \let\includegraphics\DIFaddincludegraphics} 
\DeclareRobustCommand{\DIFaddendFL}{\DIFOaddendFL \let\includegraphics\DIFOincludegraphics} 
\DeclareRobustCommand{\DIFdelbeginFL}{\DIFOdelbeginFL \let\includegraphics\DIFdelincludegraphics} 
\DeclareRobustCommand{\DIFdelendFL}{\DIFOaddendFL \let\includegraphics\DIFOincludegraphics} 

\begin{document}

 \title{Avenues to generalising Bell inequalities}
 \author{Marcin Karczewski${}^{1}$}
\address{${}^{1}$International Centre for Theory of Quantum Technologies, University of Gda\'nsk, 80-309 Gda\'nsk, Poland}
 \author{Giovanni Scala${}^{1,2}$}
\address{${}^{2}$Faculty of Physics, University of Warsaw, Pasteura 5, 02-093 Warsaw, Poland}
\address{${}^{1}$International Centre for Theory of Quantum Technologies, University of Gda\'nsk, 80-309 Gda\'nsk, Poland}
 \author{Antonio Mandarino${}^{1}$}
\address{${}^{1}$International Centre for Theory of Quantum Technologies, University of Gda\'nsk, 80-309 Gda\'nsk, Poland}
\author{Ana Bel\'en Sainz${}^{1}$}
\address{${}^{1}$International Centre for Theory of Quantum Technologies, University of Gda\'nsk, 80-309 Gda\'nsk, Poland}
\author{Marek \.Zukowski${}^{1}$}
\address{${}^{1}$International Centre for Theory of Quantum Technologies, University of Gda\'nsk, 80-309 Gda\'nsk, Poland}

\date{\today}

\begin{abstract}
  Characterizing the set of all Bell inequalities is a notably hard task. An insightful method of solving it in case of Bell correlation inequalities for scenarios with two dichotomic measurements per site  -- for arbitrary number of parties -- was given in Refs.[\textit{Phys.~Rev.~A 64, 010102(R) (2001)}] and [\textit{Phys.~Rev.~A 64, 032112 (2001)}].  
  \mk{ Using complex-valued correlation functions, we generalize their approach to a broader class of Bell scenarios, in which the parties may choose from more than 2 multi-outcome measurements.}
 Although the resulting families of Bell inequalities are not always tight, their coefficients have an intuitively understandable structure.\mk{We probe their usefulness by numerically testing their ability to detect Bell nonclassicality in simple interferometric experiments.} Moreover, we identify a similar structure in the CGLMP inequality expressed in a correlation-based form, which allows us to generalise it to three parties. 
\end{abstract}

\maketitle
\section{Introduction}
One of the main lessons we learned last century is that nature is not classical, and one particular case where this emerges is in a Bell test. The original idea of Bell to show if the prediction of quantum mechanics 
were violating the prescription of a local realistic theory (which in this manuscript we endorse as a notion of \textit{classicality}) was formalized 
choosing systems composed by two parties on which dichotomic observables are 
measured in two different settings each. 
On the experimental side, a first evidence for the violation of local realism was given in  Ref.~\cite{freedman1972experimental},  and  the irrefutable experimental  demonstration that closes loopholes,  a quite challenging task, was only recently presented \cite{hensen2015loophole,loopholefree1, loopholefree2,PhysRevLett.119.010402}. On the theoretical point, the question 
of how the bound on the quantum expectation value of the correlation function 
exceeds the maximum classical one was addressed in Ref.~\cite{Tsirelson80}. 
In parallel, some other works started investigating richer scenarios for Bell tests with $N$ parties, with $k$ measurement settings, and $d$ outcomes each \cite{Mermin-Schwarz,Mermin,BelKly93,zukowski1997realizable, PhysRevLett.85.4418,CGLMP02, WW01, ZB02}, hereafter referred to as $\mathcal{B}_{N,k, d}$. 
Finding the maximum ratio between the quantum and the classical bounds (and the set of states that realize it) for Bell tests in all but a few simplest scenarios is still an open and challenging topic, and a review of the state-of-the-art is presented in Ref.~\cite{Review_NL}. 

If one merely focuses on a particular Bell inequality in a fixed Bell scenario, then there are numerical tools with which to approximate the ratio between the maximum quantum value and the classical bound for the inequality \cite{NPA2007, NPA2008}. Such a ratio is an intuitive indicator of how robust the corresponding Bell experiment is for certifying a violation of local realistic models. However, these tools become impractical for Bell scenarios with larger values for $N$, $k$, and $d$. In particular, they  usually do not suggest a candidate quantum experiment of the desired dimensionality that would achieve such an optimal ratio.
Therefore, if one has a particular quantum system in mind to use in a Bell test, then a different approach to finding a Bell inequality that certifies the non-classicality of the statistical data becomes of value. 

In this paper, we focus on systems composed by two or more parties, which are used for performing Bell tests in a scenario $\mathcal{B}_{N,k, d}$ where $k$ and $d$ are not necessary binary variables. 
We define families of Bell inequalities building up on the approach of the authors of Refs.~\cite{WW01,PhysRevA.64.010102,ZB02} who derived all the Bell inequalities\footnote{Refs.~\cite{WW01, ZB02} even summarized these $2^{2^N}$ inequalities in each $\mathcal{B}_{N,2,2}$ scenario by means of a single non-linear inequality.} in a  $\mathcal{B}_{N,2,2}$ scenario (with arbitrary $N$). 
We explore the scope of our technique by performing numerical calculations for small Bell scenarios and computing the ratio between the quantum violation and the classical bound of the corresponding inequality.
This manuscript is organized as follows. 
In Section \ref{se:prel} we introduce the basic notions and notation on Bell scenarios, Bell experiments, and Bell inequalities, that are required to follow the manuscript. 
In Section \ref{sec:N22} we review the seminal work of Refs.~\cite{WW01,ZB02}, which fully answered the question we are interested in for the case of Bell scenarios with two dichotomic measurements per party (and arbitrary number of parties). In particular, Sec.~\ref{wwzb} present the results of Refs.~\cite{WW01,ZB02}, while Sec.~\ref{alternative} presents an alternative narrative to the results of Refs.~\cite{ZB02}, of relevance for Sec.~\ref{sec:conj}.
Section \ref{sec:conj} presents our first technique to define Bell inequalities in an arbitrary Bell scenario, inspired by the work of Refs.~\cite{WW01,ZB02}. We explain our idea in subsections \ref{ss:idea} and \ref{ss:extension}. Then, in subsection \ref{se:multiport} we present an interferometric setup that will serve as a testbed for violating the inequalities. Next, \ref{ss:k=d} and \ref{ss:k!=d} contain examples of inequalities stemming from our method, together with a discussion of their ability to detect departure from local realism. In particular, we show that the CGLMP inequality \cite{CGLMP02} in the $\mathcal{B}_{2,2,3}$ scenario can be re-derived with our method. Moreover, we show that CGLMP inequalities can also be generalized to a three party case and the resulting inequalities define some facets of the convex polytope \cite{pitowsky1991correlation} of values attainable by local realistic correlation functions. That is the new inequalities are ``tight''. \mk{This is assessed by checking the number of linearly independent deterministic local hidden variable models that achieve the classical bound, see Ref.~\cite{zukowski2006all} for details.  } Section \ref{sec:cglmp} presents our second technique to define new Bell inequalities, inspired by a rewriting of the CGLMP inequality, and some interesting numerical results. Section \ref{sec:discussion} concludes with a discussion on future research avenues.

\section{Preliminaries on Bell inequalities}\label{se:prel}
A Bell experiment is to test whether, a set of probabilities related with coincident events, a.k.a. correlations, at two or more spatially separated measurement stations,  can admit any local realistic model (essentially, a classical probabilistic model, with underlying Einsteinian Locality). This is under the proviso of randomly changing several different settings of all the local measuring devices.

For instance, in a bipartite scenario (where we refer to the parties as Alice and Bob) these correlations can be tested for an array of probabilities $\pr_{\A\Bo|\X\Y} := \{\{p(ab|xy)\}_{a\in\A\,,\,b\in\Bo}\}_{x\in\X\,,\,y\in\Y}$, where $\X$ is the set of possible values that the classical variable $x$ denoting Alice's choice of measurement setting can take, and $\A$ is the set of possible values that the classical variable $a$ denoting Alice's measurement outcome can take (in the case of Bob, these are respectively $\Y$, $y$, $\Bo$, and $b$).

Bell inequalities are the fundamental tool to certify that a set of probabilities $\pr_{\A\Bo|\X\Y}$ does not admit any joint local-hidden-variable (LHV) model. 

A linear Bell inequality involves a Bell functional, denoted by $\BFA$, that maps correlations $\pr_{\A\Bo|\X\Y}$ into real numbers.
The local realistic bound of the Bell functional, which we denote by $\beta_C$, is given by the maximum value that $\BFA(\pr_{\A\Bo|\X\Y})$ can take when evaluated over correlations $\pr_{\A\Bo|\X\Y}$ that admit an LHV model. The Bell functional together with its classical bound then define the Bell inequality
\begin{equation}\label{eq:1}
 \BFA(\pr_{\A\Bo|\X\Y}) \leq \beta_C   
\end{equation}
which all correlations that admit an LHV model satisfy. 

If a set of probabilities $\pr_{\A\Bo|\X\Y}$ yields a value larger than $\beta_C$ when the functional $\BFA$ is evaluated on it, then $\pr_{\A\Bo|\X\Y}$ is said to \textit{violate} the Bell inequality, which shows that it does not admit of an LHV model. In this manuscript, we study violations of Bell inequalities by focusing on the ratio 
\begin{equation}\label{eq:R}
    R=\frac{\BFA(\pr_{\A\Bo|\X\Y})}{\beta_C},
\end{equation}
 which when it is greater than 1 signals a violation of the inequality in Eq.~\eqref{eq:1}. Note that the ratio \eqref{eq:R} is defined only for $\beta_C\neq0$, which is not always the case. For instance, the celebrated Clauser-Horne inequality \cite{clauser1974experimental} does not meet this requirement. Nonetheless, the $R$ ratio will always be well-defined for the Bell inequalities we are going to investigate in this work. 

Therefore, a question we study is how to construct functionals $\BFA$ in an arbitrary Bell scenario $\mathcal{B}_{N,k, d}$ which are useful for detecting the non-classicality of correlations. \mk{We do not address this issue for general quantum correlations, but focus on the ones stemming from} specific experimental setups, \mk{relevant from the practical point of view.}
\mk{In particular, we propose a method of} defining a family of functionals $\BFA$, which extends the ground-breaking construction of Refs.~\cite{WW01,ZB02}. \mk{We later test their ability to detect violations of local realism in a class of interferometric setups defined in Sec.~\ref{se:multiport}. To this end, we  investigate the maximal value of the ratio R obtainable in such experiments through our Bell functionals. }

\section{Bell scenarios with two dichotomic measurements per site}\label{sec:N22}
In this section we briefly review the works by Werner and Wolf \cite{WW01} and by {\.{Z}}ukowski and Brukner \cite{ZB02}, which motivated the particular avenue we explore to generalise Bell inequalities. Moreover, we present an alternative description of the work by Ref.~\cite{ZB02}, which will underpin the results we present in the next section. 

\subsection{Full set of inequalities for the $\mathcal{B}_{N,2,2}$ scenario for correlation functions.}\label{wwzb}
If one studies the $\mathcal{B}_{N,2,2}$ scenarios, and this in terms of Bell-GHZ correlation functions (see further), as this is done in Refs.~\cite{WW01,PhysRevA.64.010102,ZB02}, then it turns out that is very easy to derive  the full set of tight inequalities for the problem, and therefore find a necessary and sufficient condition for existence of local realistic models for such scenarios. Thus the method used in the papers seems to be highly potent, and thus it is worth to study possibilities of extending this approach to different Bell scenarios.

Here we present the derivation of this set of tight Bell inequalities following the paper of \cite{WW01}.
Let the classical variable $x_j \in \{0,1\}$ denote the choice of measurement of the $j$-th party in the experiment, with $j \in \{1,\ldots,N\}$. The measurement choices of all the parties are then collected in a string that we denote by $\mathbf{x}$, where its $j$-th component corresponds to $x_j$. Given that there are two measurements for each of the $N$ parties, the string $\mathbf{x}$ can take $2^N$ possible values.  The conditional probability distribution is in this case  $\pr_{\A,\ldots,\A|\X,\ldots,\X}$, where the subscript contains $N$ times the set $\A = \{0,1\}$ and $N$ times the set $\X=\{0,1\}$ to encompass all the parties in the experiment\footnote{In this work we adopt the convention in which outcomes are labelled $0,1,\ldots,d-1$ and assigned roots of unity, $1,exp{\frac{2\pi i}{d}},\ldots,exp{\frac{2\pi(d-1) i}{d}}$, as values.}.

After specifying this notation, the first step in the formalism is to focus on the discrete Fourier transform of the conditional probability distribution $\pr_{\A,\ldots,\A|\X,\ldots,\X}$.
The particular discrete Fourier transform considered has components given by the full correlation functions 

\begin{align}\label{eq:2}
 E_{\mathbf{x}} = \sum_{\mathbf{a} \in \A^{ N}} (-1)^{a_1+\ldots+a_N} \, p(\mathbf{a} | \mathbf{x})\,,   
\end{align}
where $\mathbf{a} = (a_1,\ldots, a_N)$ is the array collecting the measurement outcomes by all the parties. 

Ref.~\cite{WW01} then normalises the coefficients of the Bell inequalities so that they take the form: 
\begin{align}\label{eq:WWZB}
    \BFA(\pr_{\A,\ldots,\A|\X,\ldots,\X}) = \sum_{\mathbf{x} \in \X^{ N}} q(\mathbf{x}) \, E_{\mathbf{x}} \leq 1\,,
\end{align}
where the classical bound is taken to be $\beta_C=1$. On this grounds, Ref.~\cite{WW01} then characterises all the possible ways one can choose the coefficients\footnote{ Notice that, in a way, the coefficients $\{q(\mathbf{x})\}_{\mathbf{x} \in \X^{ N}}$ can be thought of ways to \textit{probe} the correlations -- different choices of the coefficients will yield different tests. } $\{q(\mathbf{x})\}_{\mathbf{x} \in \X^{ N}}$ to be. They show that there are $2^{2^N}$ ways of specifying $\{q(\mathbf{x})\}_{\mathbf{x} \in \X^{ N}}$, and the different inequalities that they specify  give rise to a complete\footnote{ A set of Bell inequalities is \textit{complete} when a correlation is classical if and only if it satisfies all of the inequalities in the set.}  set of Bell inequalities that characterises precisely the set of classical correlations (i.e., those that admit of an LHV model). In a nutshell, 
\begin{align}\label{eq:4}
q(\mathbf{x}) = 2^{-N} \sum_{\mathbf{r} \in \in \X^{ N}} f(\mathbf{r}) \, (-1)^{\mathbf{r} \cdot \mathbf{x}},
\end{align}
where $f(\mathbf{r}) = \pm 1$, and $\mathbf{r}$ is an array of $N$ elements that take values 0 or 1 (just like $\mathbf{x}$). One can see that there are $2^{2^N}$ ways to choose the values that $f$ assigns to the arrays $\mathbf{r}$, which gives rise to the $2^{2^N}$ different Bell inequalities. 

\subsection{Alternative derivation}\label{alternative}
We will now present a different method of characterizing all Bell correlation inequalities in the  $\mathcal{B}_{N,2,2}$ scenario, based on the insights presented in Ref.~\cite{PhysRevA.64.010102,ZB02}. To keep things simple, we focus on the bipartite case, but the reasoning for $N>2$ is analogous.
This derivation explicitly relies on an approach in which local realistic models are formulated using tensor products of the deterministic ones for the observers, see Ref.~\cite{PhysRevA.64.010102}. Any such model is a convex combination of the deterministic ones. Identifying an orthonormal basis formed by these deterministic models allows one to derive a single non-linear Bell inequality which is equivalent to all tight Bell inequalities for the problem. Simply, one can re-express the convex combination in a form which uses only those deterministic models belonging to the basis, just like this is in any vector space. It can be easily shown that the {\it moduli} of the expansion coefficients with respect to  the basis deterministic models adds up to less than $1$ for proper local realistic models. This gives the single non-linear Bell inequality (for the given $N$). This specific feature of any local realistic model for the considered set of correlation functions,  allows a simple constructive  proof that the non-linear inequality form a necessary and sufficient condition of a local realistic description of the correlation functions, see Ref.~\cite{ZB02}. 

Here we present some details for $N=2$.
The four correlation functions from Eq.~(\ref{eq:2}), each for a different set of settings, can be put in a form of a vector 
$ E =( E_{00} ,  E_{01}, E_{10} ,  E_{11}), $
where the indices denote the local settings\footnote{Equivalently, one may think of the set of four correlation functions as forming a tensor, which can be written in matrix form $\hat{E}  = \left[ \begin{smallmatrix} E_{00} & E_{01} \\ E_{10} & E_{11}  \end{smallmatrix} \right]$.}.
As any stochastic LHV model can be reproduced by a convex mixture of deterministic ones, any LHV model of  $E$ must be of the form
\begin{equation}\label{Elhv}
E = \sum_{a_1,a_2;b_1,b_2=0,1} p(a_1,a_2,b_1,b_2) \big( (-1)^{a_1+ b_1} ,(-1)^{ a_1+ b_2},(-1)^ {a_2+ b_1} , (-1)^{ a_2+ b_2}\big), 
\end{equation}
where $a_i$ and $b_i$ denote the deterministic values $\{0,1\}$ that the LHV model assigns to the outcomes \blk of the $i$-th measurement of the first and second party, and $ p(a_1,a_2,b_1,b_2)$ are the probabilities that such deterministic outcomes appear in the given model. 
 Ref.~\cite{ZB02} aims to show that $E_{LR}$  (i.e.,  the vector $E$ of correlation functions that admit an LHV model) 
 satisfies a constraint in a form of an inequality.  To this end, in our alternative approach, we can consider the inequality
\begin{align}
    \sum_{l,n=0,1} |E\cdot \mathbf{v}_{l,n} |\leq \beta_C\,,
\end{align}
where $\beta_C$ is the classical bound of the inequality. Now, we choose a specific form for the `coefficients' $\mathbf{v}_{l,n}$: 
\begin{equation}\label{eq:vln}
   \mathbf{v}_{l,n} = v_l\otimes v_n\,, \quad l,n \in \{0,1\}\,,
\end{equation}
with
\begin{equation}
   v_l := \frac{1}{\sqrt2}(1,(-1)^l)\,,\qquad
   v_n := \frac{1}{\sqrt2}(1,(-1)^n)\,.
\end{equation}
Notice also that the vectors $\{v_l\}_{l\in \{0,1\}}$ form an orthogonal basis of $\mathbb{R}^2$, and that they correspond to a subset of the deterministic assignments of outcomes to Alice's measurements (i.e., a subset of the LHV models for Alice). The same comment holds for $\{v_n\}_{n\in \{0,1\}}$ and Bob's LHV models. 

As the scalar product of any $((-1)^{a_1},(-1)^{a_2})$  with any $v_l$ is $\pm 1,$
or $0$ -- and similarly for $((-1)^{b_1},(-1)^{b_2})$ and $v_n$ -- this Bell inequality follows:
\BE \label{WWZBmod}
\sum_{l,n=0,1} |E\cdot v_l\otimes v_n |\leq 2 \,.
\EE
As any two  real numbers $r_1$ and $r_2$ satisfy $|r_1\pm r_2|\leq
|r_1|+|r_2|$, the above nonlinear inequality is equivalent to the full set of linear ones of the form
\begin{equation}\label{WWZBset}
  \left|  \sum_{l,n=0,1} (-1)^{g(l,n)} E\cdot v_l\otimes v_n \right| \leq 2, 
\end{equation}
where $g(l,n)$ is an arbitrary function returning 0 or 1. 
Since there are 16 ways of choosing $g(l,n)$, this method allows one to construct 16 Bell inequalities. For instance, when $g(l,n)= l n$.  We get $|E_{LR}\cdot(1,1,1,-1)|\leq2$, which is exactly the famous CHSH inequality. Note that in this case $(-1)^{g(l,n)}$ does not factorize into a function of $n$ and a function of $l$. Only in such cases we get non-trivial linear Bell inequalities.

A simple constructive proof of the fact that the non-linear inequality forms a necessary and sufficient condition for a local realistic model of the involved set of values of the correlation functions can be found in Ref.~\cite{ZB02}. Thanks to the fact that just one non-linear inequality (and its obvious generalization to $N$ parties) forms the condition for local realistic description, one can also derive a general condition for $N$ qubit states to violate local realism within the considered scenario, see Ref.~\cite{ZB02}.

\section{Bell inequalities for $d>2$.}\label{sec:conj}

\subsection{Basic idea }\label{ss:idea}
We are going to generalize the approach presented in the previous section to Bell scenarios beyond two dichotomic measurements. We begin with a study of a possible generalization of the approach to $\mathcal{B}_{N,d,d}$ scenarios, that is, a situation in which the number of measurements each party can perform matches the number of possible outcomes. This is because the connection between our approach and the ones of Refs.~\cite{WW01,ZB02} is more evident in that case. A general $\mathcal{B}_{N,k,d}$ scenario will be considered later, once we explain our basic idea.\footnote{A different approach to multipartite correlation Bell inequalities, focusing on the symmetries of their coefficients, was developed in \cite{bancal,grandjean}.} 

\mk{Following Ref. \cite{zukowski1997realizable}, we are going to use (generalized) ``Bell numbers assignment''. It consists in assigning  values given by a root of unity ($\alpha_d^h$ with $h=0,1,\ldots,d-1$, where $\alpha_d=e^{\frac{2 \pi i }{d}}$)
to the measurement outcomes $\mathbf{a} = (a_1,\ldots, a_N)$  
via $a_j \, \rightarrow \, \alpha_d^{a_j}$. This way, certain properties of the $\pm1$ assignment used in the dichotomic case can be generalized, for instance $\sum_{k=1}^d\alpha^k=0$ and $|\alpha^k|=1$. The product correlation function then reads
\begin{align}\label{eq:fullcorr}
E_{\mathbf{x}} = \sum_{\mathbf{a} \in \A^{ N}} \alpha_d^{a_1}\ldots \alpha_d^{a_N} \, p(\mathbf{a} | \mathbf{x}).
\end{align}

}

\mk{One can also arrive at Eq.~\eqref{eq:fullcorr} via a different reasoning. }Notice that the correlation function of Eq.~\eqref{eq:2} used in the case of the dichotomic measurements can be thought of as an element of a Fourier transform of the probability distribution for the given set of local settings\footnote{Notice that a discrete Fourier transform of a function has the same number of values as the function. Here we consider just one of such values, i.e., $E_{\mathbf{x}}(\mathbf{r}) = \sum_{\mathbf{a} \in \A^{ N}} \alpha_d^{\mathbf{a}\cdot \mathbf{r}} \, p(\mathbf{a} | \mathbf{x})$ with $\mathbf{r}$ the vector of `all ones'. }. In order to extend this interpretation to $d$-outcome measurements, we need to use the following correlation function
$ E_{\mathbf{x}} = \sum_{\mathbf{a} \in \A^{ N}} \alpha_d^{(a_1+\ldots+a_N)} \, p(\mathbf{a} | \mathbf{x})\,,  
$
\mk{which is exactly what we have in Eq.~\eqref{eq:fullcorr}.}
Next we proceed like in Sec.~\ref{alternative}, that is: define a vector $E$ build out of all correlation functions $ E_{\mathbf{x}}$ for all considered valued of the settings $\mathbf{x}$. The form of $ E_{\mathbf{x}}$ when it may be realised by an LHV model is analogous to Eq.~\eqref{Elhv}. Now, the vectors 
\begin{equation}\label{basis}
    v_{h_1}\otimes\ldots\otimes v_{h_N} = \frac{1}{d^{\frac{N}{2}}}\left(1,\alpha_d ^{h_1},\ldots,\alpha_d^{(d-1)h_1}\right)\otimes\ldots\otimes \left(1,\alpha_d ^{h_N},\ldots,\alpha_d^{(d-1)h_N}\right),
\end{equation}
with $h_1,\ldots,h_N=0,1,\ldots,d-1$,  form an orthonormal basis for the space where the vectors of deterministic outcome-assignments used in $E_{LR}$ belong. This basis is self-conjugate, meaning that a complex scalar product (here denoted by $(\cdot , \cdot )$) between its two elements yields 
$$
(v_{h_1}\otimes\ldots\otimes v_{h_N},v_{h'_1}\otimes\ldots\otimes v_{h'_N})=\delta_{h_1,h'_1}\ldots \delta_{h_N,h'_N}. 
$$
The main idea now is to use the products $(E_{LR},v_{h_1}\otimes\ldots\otimes v_{h_N})$ to characterize correlations attainable by LHV models.  One could hope that a nonlinear inequality  which generalizes Eq.~\eqref{WWZBset}
\BE\label{WWZBgen}
\sum_{h_1,\ldots,h_N=0,1,\ldots,d-1} |(E, v_{h_1}\otimes\ldots\otimes v_{h_N}) |\leq \beta_C.
\EE
could encompass all the Bell correlation inequalities in the given scenario, similarly to the case for $\mathcal{B}_{N,2,2}$.
However, this is not true. In the general case (in contrast to the $\mathcal{B}_{N,2,2}$ scenario), most of the scalar products between vectors of deterministic results that build up $E_{LR}$ and $v_{h_1}\otimes\ldots\otimes v_{h_N}$ do not vanish. Because of that, the LHV bound $\beta_C$ in Eq.~\eqref{WWZBset} becomes too large to be violated. 
 Still, one can investigate Bell functionals defined in analogy to Eq.~\eqref{WWZBset}, given by
 
\begin{equation}\label{abs}
 \BFA_{\text{ABS}}(\pr_{\A,\ldots,\A|\X,\ldots,\X})=   \Big|\sum_{h_1,\ldots,h_N=0,1,\ldots,d-1} \alpha_d^{g(h_1,\ldots,h_N)} (E, v_{h_1}\otimes\ldots\otimes v_{h_N}) \Big|,
\end{equation}
and
\begin{equation}\label{re}
    \BFA_{\rp}(\pr_{\A,\ldots,\A|\X,\ldots,\X})= \rp\Big[\sum_{h_1,\ldots,h_N=0,1,\ldots,d-1} \alpha_d^{g(h_1,\ldots,h_N)} (E, v_{h_1}\otimes\ldots\otimes v_{h_N}) \Big],
\end{equation}
where $g$ is an arbitrary function returning $0,1,\ldots,d-1$.
For the case of $\BFA_{\rp}$ one could also investigate a more general expression $\rp\Big[z\sum_{h_1,\ldots,h_N=0,1,\ldots,d-1} \alpha_d^{g(h_1,\ldots,h_N)} (E, v_{h_1}\otimes\ldots\otimes v_{h_N}) \Big]$, where $z$ is an arbitrary complex number. However, for the considerations in this paper\footnote{We did a preliminary numerical search for bipartite scenarios with small numbers of $k$ and $d$, and noticed that most of the times a complex $z$ did not provide advantages when there's freedom on the function $g$.} we take $z=1$.
Notice that for $\BFA_{\text{ABS}}$ this global factor $z$ does not bring in any extra freedom, since it just provides a re-scaling of Eq.~\eqref{abs}.
To find the classical bounds $\beta_C$ for these functionals, it suffices to look for their  maximal value on the probability distributions stemming from deterministic local hidden variable models, and take the largest one of those values. In contrast to the $\mathcal{B}_{n,2,2}$ scenario, these bounds will not be the same for all the possible choices of the $g$ function. We will be most interested in the $g$ functions that lead to inequalities specifying a facet of the classical \textit{polytope of  full-correlation functions}  (that of vectors of values of correlation functions $(E_{11}, \ldots, E_{kk})$ consistent with local realism). We refer to such inequalities as ``tight''.

\subsection{Extension to arbitrary $\mathcal{B}_{N,k,d}$ scenarios}\label{ss:extension}
Having presented our basic idea, we can now investigate what happens if the number of measurements per party $k$ is not equal to the number of possible outcomes $d$. In a nutshell, it will not always be possible to select a subset of normalized vectors of deterministic outcomes that would form their orthonormal basis, as we have done in Eq.~\eqref{basis}. For instance, if $k=2$ and $d=3$, the possible LHV outcomes a party can register form vectors $(\alpha_3^{h_1},\alpha_3^{h_2})$ with $h_1,h_2=0,1,2$. A pair of such vectors cannot be orthogonal, as $\big((\alpha_3^{h_1},\alpha_3^{h_2}),(\alpha_3^{h'_1},\alpha_3^{h'_2})\big)={\alpha_3}^{h'_1 - h_1}+{\alpha_3}^{h'_2 - h_2}\neq 0$. Clearly, if a basis is not orthogonal, it cannot be self-conjugate. Hence, for Bell scenarios with $k\neq d$ we will, in general, need to specify two different bases: one for the vectors of deterministic results, and the other conjugate to it. Aside from this caveat, our approach works just as in the $k=d$ case, leading to Bell functionals of the form in Eqs.~\eqref{abs} and \eqref{re}.

In this work we investigate the applicability of our methods to the $k\neq d$ scenarios for $k=2$. In that case, we will usually construct the basis of deterministic outcomes from the vectors $(1,1)$ and $(1,\alpha_d^{d-1})$, and the conjugate basis from $\frac{1}{1-\alpha_d}(1,-1)$ and $\frac{1}{1-\alpha_d}(-\alpha_d,1)$. This choice is not unique, as one could take any basis of deterministic outcomes as the starting point. However, as we will see in Section \ref{CGLMPconj}, it allows us to reproduce the CGLMP inequality in the $\mathcal{B}_{2,2,3}$ scenario.

\subsection{Experimental scheme}\label{se:multiport}

In this section we present an interferometric measurement scheme that one can use to find experimental (or thought-experimental) violations of Bell inequalities. Here we discuss the setup, and in the next subsections we show how it provides violations of the Bell inequalities that we construct from Eqs.~\eqref{re} and \eqref{abs}. 

\begin{figure}[t]
	\centering
\includegraphics[width= 1
\columnwidth]{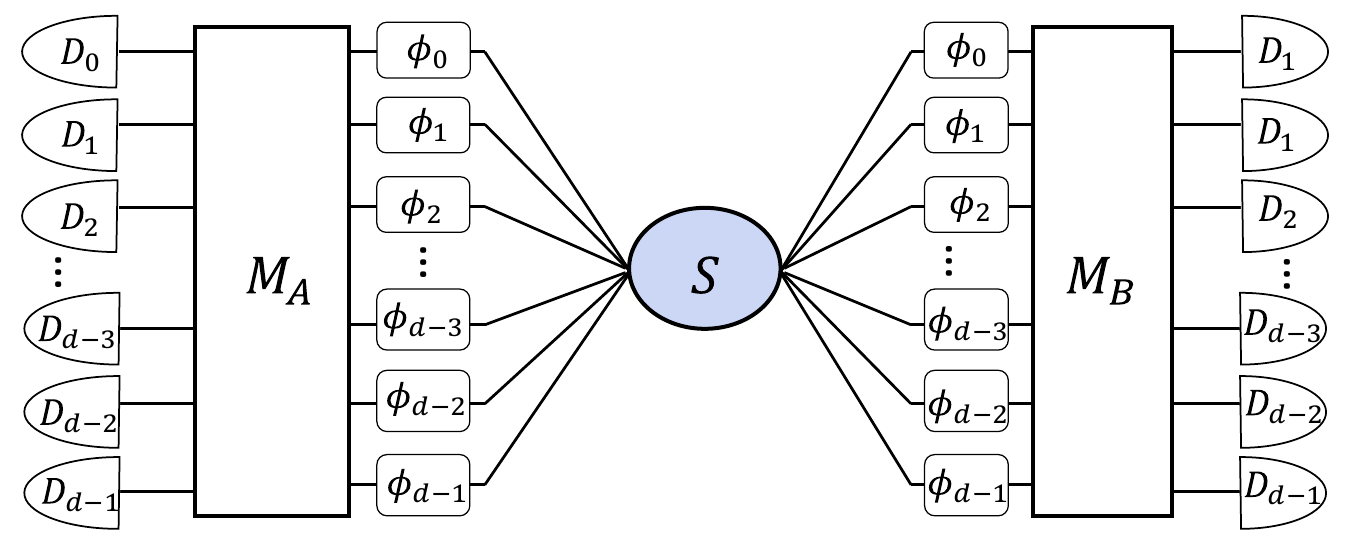}
	\caption{\label{setup}
		Interferometric scheme used to define measurements in a scenario with two parties, $d$ outcomes per measurements and an initial state of two qudits. Each party chooses a set of local phases $\{\phi_i\}$ applied in the input ports of their symmetric $d$-port. The single-photon local parts of the initial state, distributed by a source S, are encoded in the input modes, i.e., $|j\rangle$  corresponds to a single photon in $j$-th port. The index of the detector $D_i$ that registered the photon upon passing through the multiport is the outcome of the measurement.     }
\end{figure}
The photonic experimental scheme (see Fig.\ref{setup}) was put forward in Ref.~\cite{zukowski1997realizable} and later studied, for instance, in the context of multipartite GHZ correlations \cite{kaszlikowski2002greenberger}. For simplicity we present and depict the case of two parties, but the generalisation of the setup for multiple parties is straightforward. In this bipartite scheme, a source distributes a pair of $d$-dimensional photons to Alice and Bob. Each party has access to a multiport $\mathrm{M}$ with $d$ input ports and $d$ output ports.
Each input port is associated to a particular state of the input photon in the computational basis: if the photon enters through port number $t$ then its state in the computational basis reads  $\ket{t}$ (note that the input and output ports are labelled from $0$ to $d-1$). 
For example, in the case $d=3$, the matrix representation in the computational basis of the unitary performed by $\mathrm{M}$ is given by
\begin{equation}\label{eq:themultiport}
\frac{1}{\sqrt3}
\begin{bmatrix}
1 & 1 & 1 \\
1 & \alpha_3 & \alpha_3^2 \\
1 & \alpha_3^2 & \alpha_3
\end{bmatrix}
\end{equation}
where $\alpha_3 = e^{i \frac{2\pi}{3}}$. That is, the multiport implements an optical Fourier transform.

Each port in the multiport is subject to a phase-shift operation before it enters the multiport $\mathrm{M}$, denoted  $U_{PS}^{(x)}$ for Alice and $ U_{PS}^{(y)}$) for Bob,  and this is what gives each party the freedom to choose and specify ``measurement settings''  (hence why those unitaries depend on the classical variables $x$ and $y$). The idea is a follows: let $(\phi_0,\ldots,\phi_{d-1})$ denote the collection of phases for each of the ports, where $\phi_j$ denotes the phase-shift in port $j$. Then, each specification of the set $(\phi_0,\ldots,\phi_{d-1})$ corresponds to a different measurement choices. In other words, the measurement choice for Alice labeled by $x$ is associated to a choice of phases $(\phi^{(x)}_0,\ldots,\phi^{(x)}_{d-1})$ in the unitary phases shift operator $U_{PS}^{(x)}$ applied to the input system. 

Now, the `outcome' of the experiment will correspond to the output port in which the photon is detected. For instance, if Alice detects the photon exiting through port $t$, then the outcome of the experiment is labelled by $a=t-1$. We see then that the number of input/output ports that our multiport $\mathrm{M}$ must have is given by the number of outcomes in the Bell experiment. 

We will now present the explicit form that the full-correlation functions $E_{xy}$ from Eq.~\eqref{eq:fullcorr} takes in this specific setup, taking $d=3$.  
These full-correlations are computed for an input state of the form
\begin{equation}
\ket{\Psi} = \sum_{j,\ell=0}^{2} s_{j\ell} \ket{j\ell}\,,
\end{equation}
with $\sum_{j,\ell=0}^{2} |s_{j\ell}|^2=1$. 
Let $|\Psi'\rangle$ denote the final state of the system once it's been through the phase shifter and the multiport. Born's rule then sets $p(ab|xy)=|\langle ab|\Psi'\rangle|^2$. 
The state $\ket{\Psi'}$ arises from $\ket{\Psi}$ as follows: 
\begin{align}
    \ket{\Psi'} &= \left(\mathrm{M}_A \otimes \mathrm{M}_B\right)\left(  U_{PS}^{(x)}\otimes U_{PS}^{(y)}\right)  \ket{\Psi} = \mathrm{M}_A \otimes \mathrm{M}_B \sum_{j,\ell=0}^{2} s_{j\ell} e^{i\phi^{(x)}_j} e^{i\varphi^{(y)}_{\ell}} \ket{j\ell} \,,
\end{align}
where $\left(\phi^{(x)}_0,\ldots,\phi^{(x)}_{d-1}\right)$ is the collection of phases associated to Alice's measurement $x$, and $(\varphi^{(y)}_0,\ldots,\varphi^{(y)}_{d-1})$ is the collection of phases associated to Bob's measurement $y$.

Using the form of $\mathrm{M}_A$ and $\mathrm{M}_B$ given in Eq.~\eqref{eq:themultiport}, it follows that
\begin{align}
    \ket{\Psi'} &= \mathrm{M}_A \otimes \mathrm{M}_B \sum_{j,\ell=0}^{2} s_{j\ell} \, e^{i\phi^{(x)}_j} e^{i\varphi^{(y)}_{\ell}} \ket{j\ell} \\
    &= \frac{1}{3} \sum_{u,v=0}^2 \sum_{j,\ell=0}^{2}  e^{i\phi^{(x)}_j} e^{i\varphi^{(y)}_{\ell}} \, \alpha^{u j} \alpha^{v \ell} \, s_{j\ell} \,  \ket{uv} \,.
\end{align}
From there we obtain
\begin{align}
    \langle ab|\Psi'\rangle = \frac{1}{3} \sum_{j,\ell=0}^{2}  e^{i\phi^{(x)}_j} e^{i\varphi^{(y)}_{\ell}} \, \alpha^{a j} \alpha^{b \ell} \, s_{j\ell} \,.
\end{align}
The expression for the full-correlation functions $E_{xy}$ turns out to be
\begin{align}
    E_{xy} &= \sum_{a,b=0}^2 \alpha_3^{a+b} p(ab|xy)  \nonumber \\
    &= \sum_{a,b=0}^2 \alpha_3^{a+b} \frac{1}{9} \left( \sum_{j,\ell=0}^{2}  e^{i\phi^{(x)}_j} e^{i\varphi^{(y)}_{\ell}} \, \alpha_3^{a j} \alpha_3^{b \ell} \, s_{j\ell} \right) \left( \sum_{j',\ell'=0}^{2}  e^{-i\phi^{(x)}_{j'}} e^{-i\varphi^{(y)}_{\ell'}} \, \alpha_3^{-a j'} \alpha_3^{-b \ell'} \, s_{j'\ell'}^* \right) \nonumber \\
    &= \frac{1}{9} \sum_{j,\ell=0}^{2} \sum_{j',\ell'=0}^{2} e^{i (\phi^{(x)}_j - \phi^{(x)}_{j'} + \varphi^{(y)}_{\ell} - \varphi^{(y)}_{\ell'})} \, s_{j\ell} \, s_{j'\ell'}^* \, \left( \sum_{a=0}^2 \alpha_3^{a(j-j'+1)} \right) \left( \sum_{b=0}^2 \alpha_3^{b(\ell-\ell'+1)} \right)\,.
\end{align}
Now notice that since $\alpha_3$ is a third root of unity and the sum runs through the three possible values of $a$ it yields

\begin{align*}
    \sum_{a=0}^2 \alpha_3^{a(j-j'+1)} = \begin{cases}
        3 \quad \text{if} \quad j-j'+1=0 \,, \\
        0 \quad \text{otherwise},
    \end{cases}
\end{align*}

and similarly for $\sum_{b=0}^2 \alpha^{b(\ell-\ell'+1)}$. 

Hence, the full-correlation functions can be written as
\begin{align}
    E_{xy} &= \sum_{j,\ell=0}^{2} e^{i (\phi^{(x)}_j - \phi^{(x)}_{j+1} + \varphi^{(y)}_{\ell} - \varphi^{(y)}_{\ell+1})} \, s_{j\ell} \, s_{j+1,\ell+1}^*\,,
\end{align}
where the sums $j+1$ and $\ell+1$ are taken mod 3. 

\subsection{Application to the $\mathcal{B}_{2,3,3}$ Bell scenario}\label{ss:k=d}

The expressions $\BFA_{\text{ABS}}$ and $\BFA_{\rp}$ from Eqs.~\eqref{abs} and \eqref{re}, respectively, define large families of Bell inequalities. To probe their usefulness, in this subsection we focus on a special case in which the $g$ function is symmetric, i.e., $g((x,y))=g((y,x))$\mk{, and the correlations stem from the experimental setups defined in Sec.~\ref{se:multiport}}. We considered a Bell scenario with 3 measurements per party (i.e., $\X=\{0,1,2\}=\Y$) and three outcomes per measurement (i.e., $\A=\{0,1,2\} = \Bo$),
 and optimized the $R$ ratio over all symmetric $g$ functions for both $\BFA_{\text{ABS}}$ and $\BFA_{\rp}$. The possibility of violations of inequalities were investigated in the interferometric setup described in Sec.~\ref{se:multiport} by optimising over the choice of measurement settings for Alice and Bob given by the local phases, and the input state $\ket{\Psi}$. Unfortunately, the inequalities that we found are not tight, i.e. they do not define a facet of a local realistic polytope  of full-correlation functions. 

In case of the inequalities based on norm,  $\BFA_{\text{ABS}}$, the largest value of the ratio we found was $R\approx1.0482$ for the function $ g(x,y)=\delta_{x,2}\delta_{y,2}$.

Regarding the inequalities based on taking the real part of expression, $\BFA_{\rp}$, the best ratio of violation we found is $R\approx1.167$ for  
 $g(x,y)=(1-\delta_{x,0})(1-\delta_{y,0})$.
It is worth mentioning that this ratio is larger than the one corresponding to an inequality derived by \.Alsina et al.~\cite{Zyczkowski2016operational} ($R\approx1.137$), who considered the same Bell scenario. 

The maximum violation ($R\approx1.167$) is attained for the state 
\begin{align} \label{psi}
    \ket{\Psi}_{MAX}=a\ket{00}+b\ket{11}+a\ket{22},
\end{align}
where $a\approx-0.627,\,b\approx0.462$. Although the structure of this state looks very similar to that of the  state that maximally violates the CGLMP inequality \cite{CGLMP02}, its $\frac{a}{b}$ ratio is different from that of the latter.

\section{Application to $\mathcal{B}_{n,2,d}$ Bell scenarios}\label{ss:k!=d}

We now turn our attention to the case when the number of measurements each party can make does not match the number of their potential outcomes. As we have seen in Sec.~\ref{ss:extension}, in that case there is no orthogonal basis formed out of a subset of local deterministic models that we could use to construct our inequalities. Facing this, we are going to use the bases given by the vectors $\frac{1}{1-\alpha_d}(1,-1)$ and $\frac{1}{1-\alpha_d}(-\alpha_d,1)$. They form a conjugate basis with respect to the non-orthogonal basis formed  by the vectors  $(1,\alpha^*_d)=(1,\alpha^{d-1}_d)$ and $(1,1)$. Note that the latter one is formed out of possible deterministic local models, as it was in the case of the approach of the previous section.  To probe the effectiveness of our approach, we have checked numerically if inequalities based on norm ($\BFA_{\text{ABS}}$) and real part ($\BFA_{\rp}$) can be violated in different Bell scenarios with two measurement choices per party, when the $g$ function is a simple product one, $g(k,l,m)= k l m$ for $N=3$. The results are summarised in Tab.~\ref{tab:res_gs}. In most scenarios the inequalities $\BFA_{\rp}$ proved to be better than $\BFA_{\text{ABS}}$  at certifying significant violations of local realism.  The ratios grow with the number of parties, but no clear pattern is visible with respect to the number of outcomes.

\begin{table}\label{tab:res_gs}
\begin{centering}
\begin{tabular}{|c|c|c|}
\hline 
$\left(N,k,d\right)$ & $R_\mathrm{Re}$ & $R_\mathrm{ABS}$\tabularnewline
\hline 
\hline
 (2,2,2) & 1.41421 & 1.41421 \\
 (2,2,3) & 1. & 1. \\
 (2,2,4) & 1.03344 & 1. \\
 (2,2,5) & 1.40564 & 1.44544 \\
 (2,2,6) & 2. & 1.71638 \\
 (2,2,7) & 4.28896 & 2.1611 \\
 (2,2,8) & 4.05497 & 2.54065 \\
 (2,2,9) & 4.41147 & 2.97816 \\
 (2,2,10) & 4.76743 & 3.42451 \\
 (2,2,11) & 5.16095 & 3.85207 \\
 (2,2,12) & 5.78109 & 4.35003 \\
 \hline 
\end{tabular}
\quad
\begin{tabular}{|c|c|c|}
\hline 
$\left(N,k,d\right)$ & $R_\mathrm{Re}$ & $R_\mathrm{ABS}$\tabularnewline
\hline 
\hline 
(2,2,13) & 6.22541 & 4.77022 \\
 (2,2,14) & 6.84526 & 5.31234 \\
 (3,2,2) & 1.66667 & 1.66667 \\
 (3,2,3) & 1. & 1. \\
 (3,2,4) & 3.69497 & 1.12111 \\
 (3,2,5) & 3.78029 & 1.93921 \\
 (4,2,2) & 1.84277 & 1.84277 \\
 (4,2,3) & 1. & 1. \\
 (4,2,4) & 4.19176 & 1.50835 \\
 (5,2,2) & 1.97456 & 1.97456 \\
 (5,2,3) & 1. & 1.79252 \\
\hline 
\end{tabular}
\par\end{centering}
\caption{Ratios $R_{\rp}$ and $R_{\text{ABS}}$ between maximal quantum value and local realistic bound for inequalities $\BFA_{\text{ABS}}$ and $\BFA_{\rp}$ in a Bell scenario with $N$ parties, $k=2$ measurements per party and $d$ outcomes per measurement for a total amount of 22 different scenarios. The values are obtained numerically by considering a specific interferometric setup, see Sec.~\ref{se:multiport}. The details concerning the states and the local settings related to the ratios are available in our Github repository \cite{gsgithublib}.}

\end{table}

\subsection{CGLMP inequality}\label{CGLMPconj}
In this section we will find the conjugate basis form of the CGLMP inequality  \cite{CGLMP02} in the $\mathcal{B}_{2,2,3}$ scenario and present its generalisation to more parties. The original inequality reads
\begin{align}\label{CGLMP} 
P(a=b|11)+P(a=b-1|21)+P(a=b|22) \nonumber \\+P(a=b|12)-P(a=b-1|11)-P(a=b|21)\nonumber \\
-P(a=b-1|22)-P(a=b+1|12)\leq 2.
\end{align}
The probabilities in Eq.~\eqref{CGLMP} can be expressed in terms of the correlation functions
\begin{align}
E_{xy}=P(a=b|xy)&+\alpha_3 P(a=b+1|xy)+\alpha_3^2P(a=b-1|xy),
\end{align}
where $\alpha_3=e^{\frac{2\pi i}{3}}$, and their corresponding complex conjugates $E^*_{xy}$, for instance
\begin{align}
    P(a=b|xy)=\frac{1}{3}(1+E_{xy}+E_{xy}^*).
\end{align}
Because of that, the CGLMP inequality from Eq.~\eqref{CGLMP} can be rewritten as
\BE \label{CGLMP-E}
\rp[(1-\alpha_3)E_{11}-(1-\alpha_3)E_{21}+(1-\alpha_3)E_{22}+(1-\alpha_3^2)E_{12}]\leq 3.
\EE
Note that the vector of coefficients of this inequality can be expressed as
\begin{eqnarray} \label{CGLMPexp}
(1-\alpha,1-\alpha^2,\alpha-1,1-\alpha)=3[v_1\otimes v_1+v_1\otimes v_2+ \alpha v_2
\otimes
v_1
+ \alpha^2 v_2
\otimes v_2]
\end{eqnarray}
where the two vectors $v_1= \frac{1}{1- \alpha} (-\alpha,1) $
and $v_2= \frac{1}{1- \alpha} (1,-1) $
form a conjugate basis  for the original basis $w_1=(1,1)$ and $ w_2=(1,\alpha^2)$, with respect to a complex scalar product.
This means that $(w_r,v_s)=\delta_{rs}$.
 
Having  thus  found a relevant conjugate basis, we may now try to generalise the original  CGLMP inequality to three parties. To do that, we investigate inequalities of the form
\begin{equation}\label{TENSOR-IDEA}
\rp\big[\sum_{k,l,m=1}^2 \alpha^{g(k,l,m)} (E,v_k\otimes v_l \otimes v_m)\big]\leq \beta_C
\end{equation}
where 
the function $g(k,l,m)$ has integer values. Interestingly, we found out that some inequalities of form given by Eq.~\eqref{TENSOR-IDEA} are tight, i.e., define a facet of the polytope of values of  full-correlation functions compatible with the assumptions of locality and realism.  These inequalities can be grouped into three sets, corresponding to $R$ ratios of $R_1\approx1.686$, $R_1\approx1.598$ and $R_1\approx1.457$. They can be obtained, for instance, with $g_1(k,l,m)=\delta_{k,2}\delta_{l,1}\delta_{m,2}$, $g_3(k,l,m)=\delta_{k,1}\delta_{l,2}\delta_{m,1}+2\delta_{k,2}\delta_{l,1}+2\delta_{k,2}\delta_{l,2}\delta_{m,1}$ and $g_3(k,l,m)=\delta_{k,2}\delta_{l,2}$ respectively. It is worth noting that (in the $\mathcal{B}_{3,2,3}$ scenario) Ac\'in et al.~\cite{acin2004coincidence} also provide a tight Bell inequality with $R\approx1.457$, while Alsina et al.~\cite{Zyczkowski2016operational} propose one with $R=1.688$. The relationship between these inequalities  is the subject of future work. 

\section{Generalisation of the CGLMP inequality for $d=3$ to three parties using "Bell numbers" value assignment}\label{sec:cglmp}

In this section  we present a generalization of the CGLMP inequality  to more parties which rests on the concept of Bell numbers  (in the sense of `powers of the complex number given by the $d$-th root of unity')  \cite{zukowski1997realizable} and intuitions presented in the original CGLMP paper \cite{CGLMP02}. 
Our starting point is now the fact that the original CGLMP inequality (see Eq.~\eqref{CGLMP}) can be rewritten as
\begin{align} \label{CGLMP-E-2}
\Re[(1-\alpha)E_{11}+ \alpha^2(1-\alpha)E^*_{21}+(1-\alpha)E_{22}\nonumber\\+(1-\alpha^2)E_{12}] \leq 3,
\end{align}
where $E^*_{21}$ denotes the complex conjugate of $E_{21}$.
As the above needs to hold for deterministic local hidden variable models, for any numbers $a_1,a_2, b_1, b_2$ which are powers of $\alpha$ we have
\begin{align} \label{CGLMP-E-numbers}
\Re[(1-\alpha)a_1b_1+ \alpha^2(1-\alpha)a^*_2b^*_1+(1-\alpha)a_2b_2\nonumber\\+(1-\alpha^2)a_1b_2]\leq 3.
\end{align}
Note that if we set $e_{ij}=a_ib_j$, one has $e_{11}e^*_{21}e_{22}=e_{12} $.  Remarkably,  the  three numbers  in the product are arbitrary, but $e_{12}$ is determined by them. This suggests a method of generalizing the CGLMP inequality to 3 parties. In that case the correlation functions are given by $E_{ijk}= \avg{a_ib_jc_k}$. As $(a_1b^*_2c_2)(a^*_2b^*_1c^*_2)(a_2b_2c_1)=a_1b^*_1c_1$, we can write
\begin{align} \label{CGLMP-E3}
\BFA_{\text{323}}(\pr_{\A,\A,\A|\X,\X,\X})=\nonumber\\ \Re[(1-\alpha)E_{12^*2}+\alpha^2(1-\alpha)E_{2^*1^*2^*}\nonumber\\+(1-\alpha)E_{221}+(1-\alpha^2)E_{11^*1}]\leq 3.
\end{align}

Numerical optimization, conducted for the case of multiport-based measurement described in Sec.~\ref{se:multiport}, revealed some interesting features of the inequality from Eq.~\eqref{CGLMP-E3}. Its maximal violation, $\BFA_{\text{323}}\approx4.543$, is obtained for the state
\begin{align}
\label{psi_cglmp}
    \ket{\psi}_{max}=a\ket{010}+b\ket{020}\nonumber\\+c\ket{101}+d\ket{121}+d\ket{202}+e\ket{212}\nonumber\\+b\ket{000}+e\ket{111}+c\ket{222},
\end{align}
where $a\approx0.313,\,b\approx0.299,\,c\approx0.515,\,d\approx0.035,\,e\approx0.309$. This violation is substantial, with $R\approx1.514$ (comparable to that from  Ref.~\cite{Zyczkowski2016operational}, where a different construction stemming from the CGLMP operator yielded $R\approx1.686$).
On the other hand, it is not violated by the states of the form 
\begin{align}
\label{ghz_like}
\ket{\psi}_{abc}=a\ket{000}+b\ket{111}+c\ket{222},
\end{align}
where $|a|^2+|b|^2+|c|^2=1$. 
Note that the generalized Schmidt decomposition of three particle pure states, which is based on the method presented in Ref.~\cite{Peres95}, does not have the form of Eq.~\eqref{ghz_like}, see Ref.~\cite{Sudbery}.

\section{Discussion}\label{sec:discussion} 

We have presented new ways of deriving Bell inequalities. The first of them was inspired by the papers of Refs.~\cite{WW01,ZB02} which put forward a method of constructing all Bell inequalities for scenarios with two dichotomic measurements per party.  The inequalities of Refs.~\cite{PhysRevA.64.010102,WW01,ZB02} can be expressed in terms of scalar products between vectors of values of correlation functions and  some particular basis' vectors. Importantly, this basis is conjugate to a basis formed by certain correlation vectors stemming from deterministic local realistic models. These observations allowed us to generalize the approach of Refs.~\cite{WW01,ZB02} to arbitrary $\mathcal{B}_{N,k,d}$ scenarios. Although our method does not characterize all Bell inequalities in a given scenario, it can be used to derive some of them. For instance, we have used our approach not only to re-derive the CGLMP inequality for measurements with three outcomes, but also to generalize it to three parties.

We have also proposed a different way of generalizing the three-outcome CGLMP inequality. We noticed that, in case of a local realistic description of a $\mathcal{B}_{2,2,3}$ scenario, specifying three values of correlations for deterministic models for different settings determines the value of the remaining one.  A similar structure can also be found in the $\mathcal{B}_{3,2,3}$ case, which enable us to generalize the bipartite CGLMP inequality to three parties.  

Our results suggest a plethora of problems for further study. Firstly, it would be interesting to find out how the two ways of generalizing CGLMP inequality, given by Eqs.~\eqref{TENSOR-IDEA} and \eqref{CGLMP-E3},  bring new insights  in case of more parties. Another idea would be to test them in scenarios with more than three outcomes per measurement. Next, inequalities in the scenarios with $k=d$ (the number of measurement choices matching the number of outcomes) should be studied for a broader class of $g$ functions. This case is special, as the basis of the local realistic correlation vectors is self-conjugate. Thus, one could expect that our method based on such bases could lead to some new insights. One could also check which of the known tight Bell inequalities could be obtained using the conjugate basis method. This could lead to their generalisation to more parties, as we have shown in case of the CGLMP inequality. Finally, it would be interesting to apply the conjugate basis idea to the construction of entanglement witnesses and to compare the results with a tensor-besed approach recently developed in Refs.~\cite{sarbicki2020family, Sarbicki_2020_2}.

\section*{Acknowledgments}

We acknowledge support by the Foundation for Polish Science (IRAP project, ICTQT, contract no.2018/MAB/5, co-financed by EU within Smart Growth Operational Programme). GS is thankful to Gianluca Turco for the fruitful discussions and support concerning the machine learning approach for the maximization of the correlator functions.

\appendix

\bibliographystyle{unsrt}
\bibliography{biblio.bib}

\end{document}